\documentclass[twocolumn,showpacs,aps,prl,superscriptaddress]{revtex4}
\usepackage{graphicx}
\usepackage{dcolumn}
\usepackage{amsmath}
\usepackage{epsfig}
\usepackage{multirow}
\usepackage{verbatim}
\usepackage{overpic}
\input{babarsym.tex}

\def\nonetac{\ensuremath{\slash \kern-5.5pt \etac}}
\def\nonetacsub{\ensuremath{\slash \kern-4.7pt \etac}}

\def\cb{\ensuremath{\rm CB}}
\def\epb{\ensuremath{\etac\rm B}}
\def\xpb{\ensuremath{X\rm B}}

\def\E{\ensuremath{{\cal H}}}
\def\X{\ensuremath{{\cal X}}}
\def\P{\ensuremath{{\cal P}}}
\def\C{\ensuremath{{\cal C}}}
\def\W{\ensuremath{{\cal W}}}
\def\R{\ensuremath{{\cal R}}}

\def\gevccsq{\ensuremath{{\mathrm{\,Ge\kern -0.1em V^2\!/}c^4}}\xspace}
\def\gevmass{\ensuremath{{\mathrm{\,Ge\kern -0.1em V\!/}c^2}}\xspace}

\def\ptx{\ensuremath{p^T_X}}
\def\ptetac{\ensuremath{p^T_{\etac}}}
\def\phsp{\ensuremath{\Phi}}
\def\cosx{\ensuremath{\cos\theta_{X}}}
\def\cosetac{\ensuremath{\cos\theta_{\etac}}}
\def\cosks{\ensuremath{\cos\alpha_{\KS}}}
\def\evis{\ensuremath{E_{vis}}}
\def\Ex{\ensuremath{E_{\rm ex}}}
\def\etac{\ensuremath{\eta_c}}
\def\etacp{\ensuremath{\eta_c(2S)}}
\def\psip{\ensuremath{\psi(2S)}}
\def\ecpp{\ensuremath{\eta_c\pi^+\pi^-}}

\def\kskp{\ensuremath{\KS K^+\pi^-}}
\def\mmisssq{\ensuremath{m^2_{\rm miss}}}
\def\gg{\ensuremath{\gamma\gamma}}
\def\metac{\ensuremath{m(\kskp)}}
\def\metaceq{\ensuremath{m_{3}}}
\def\mx{\ensuremath{m(\kskp \pi^+\pi^-)}}
\def\mxeq{\ensuremath{m_{5}}}
\def\beq{\begin{equation}}
\def\eeq{\end{equation}}
\def\beqa{\begin{eqnarray}}
\def\eeqa{\end{eqnarray}}

\newcommand{\BABARPubYear}    {12}
\newcommand{\BABARPubNumber}  {004}
\newcommand{\SLACPubNumber} {15052}

\def\figurebox#1#2#3{
    \def\arg{#3}
    \ifx\arg\empty
    {\hfill\vbox{\hsize#2\hrule\hbox to #2{\vrule\hfill\vbox to #1{\hsize#2\vfill}\vrule}\hrule}\hfill}
    \else
    {\hfill\epsfbox{#3}\hfill}
    \fi}

\begin{document}

\begin{flushleft}
\babar-PUB-\BABARPubYear/\BABARPubNumber\\
SLAC-PUB-\SLACPubNumber\\
\end{flushleft}

\title{\large \bf Search for resonances decaying to \boldmath $\eta_c \pi^+\pi^-$ in two-photon interactions}
 
%
\author{J.~P.~Lees}
\author{V.~Poireau}
\author{V.~Tisserand}
\affiliation{Laboratoire d'Annecy-le-Vieux de Physique des Particules (LAPP), Universit\'e de Savoie, CNRS/IN2P3,  F-74941 Annecy-Le-Vieux, France}
\author{J.~Garra~Tico}
\author{E.~Grauges}
\affiliation{Universitat de Barcelona, Facultat de Fisica, Departament ECM, E-08028 Barcelona, Spain }
\author{A.~Palano$^{ab}$ }
\affiliation{INFN Sezione di Bari$^{a}$; Dipartimento di Fisica, Universit\`a di Bari$^{b}$, I-70126 Bari, Italy }
\author{G.~Eigen}
\author{B.~Stugu}
\affiliation{University of Bergen, Institute of Physics, N-5007 Bergen, Norway }
\author{D.~N.~Brown}
\author{L.~T.~Kerth}
\author{Yu.~G.~Kolomensky}
\author{G.~Lynch}
\affiliation{Lawrence Berkeley National Laboratory and University of California, Berkeley, California 94720, USA }
% ...
\collaboration{The \babar\ Collaboration}
\noaffiliation

\begin{abstract}

We report a study of the process $\gamma\gamma \to X \to \eta_{c} \pi^{+} \pi^{-}$, where
$X$ stands for one of the resonances $\chi_{c2}(1P)$, $\eta_{c}(2S)$,
$X(3872)$, $X(3915)$, or $\chi_{c2}(2P)$. 
The analysis is performed with a data
sample of 473.9~{\ensuremath{\mbox{\,fb}^{-1}}\xspace} collected with the \babar\ detector at the
PEP-II asymmetric-energy electron-positron collider.  
We do not observe a significant signal for any channel, and calculate 90\%
confidence-level upper limits on the products of branching fractions and two-photon
widths $\Gamma_{X\to\gamma\gamma}{{\ensuremath{\cal B}\xspace}}(X\to\eta_{c}\pi^+\pi^-)$:
$15.7~\ev$ for $\chi_{c2}(1P)$,
$133~\ev$ for $\eta_c(2S)$,
$11.1~\ev$ for $X(3872)$ (assuming it to be a spin-2 state),
$16~\ev$ for $X(3915)$ (assuming it to be a spin-2 state), and
$19~\ev$ for $\chi_{c2}(2P)$.
We also report  upper limits on the ratios of branching fractions
${\cal B}(\eta_{c}(2S)\to \eta_{c} \pi^{+} \pi^{-}) / {\cal B}(\eta_{c}(2S)\to \KS K^+ \pi^-) < 10.0$ and
${\cal B}(\chi_{c2}(1P)\to \eta_{c} \pi^{+} \pi^{-}) / {\cal B}(\chi_{c2}(1P)\to \KS K^+ \pi^-) <32.9$ 
at the 90\% confidence level.

\end{abstract}


\maketitle

\def\bfDbar    {\kern 0.2em\overline{\kern -0.2em D}{}\xspace}
\def\bfDzb     {\ensuremath{\bfDbar^0}\xspace}
\def\bfDstarz  {\ensuremath{D^{*0}}\xspace}
\def\bfDzDzb   {\ensuremath{\bfDzb {\kern -0.16em \bfDstarz}}\xspace}


Two-photon fusion events provide a useful production mode to study charmonium
states with quantum numbers $J^{PC}=0^{\pm+}$, $2^{\pm+}$, $4^{\pm+}$, ...,
$3^{++}$, $5^{++}$, ...~\cite{delAmoSanchez:2011bt,GGtoZtoDD}. Dipion
transitions among these states have been experimentally studied only
in one case~\cite{:2009vg}, in contrast to the narrower vector states,
where dipion transitions have been studied extensively.
In particular, the transition amplitude for 
$\etacp\to\etac\pi^+\pi^-$~\cite{etacdef} is
expected~\cite{Voloshin:2002xh} to have the same approximately linear
dependence on the invariant-mass-squared of the dipion system as
the $\psip\to J/\psi\pi^+\pi^-$ decay~\cite{Bes:jpsipipi}. 
Phase-space integration of the squared amplitude, 
evaluated for the peak masses $M_{\etac}$ and $M_{\etacp}$~\cite{delAmoSanchez:2011bt}
of the $\etac$ and $\etacp$, respectively, 
yields $\Gamma(\etacp\to\etac\pi^+\pi^-)/\Gamma(\psip\to
J/\psi\pi^+\pi^-)\approx 2.9$.  This leads to the branching fraction
prediction $\BR(\etacp\to\etac\pi^+\pi^-) = (2.2^{+1.6}_{-0.6})\%$, where
the uncertainty is due to the uncertainty on the width of the
$\etacp$~\cite{delAmoSanchez:2011bt}. This decay may be further
suppressed due to the contribution of the chromo-magnetic interaction
to the decay amplitude~\cite{Voloshin:2006ce}.

In recent years, experiments have reported evidence for
charmonium-like states, such as the $X(3872)$~\cite{3872} and 
$Y(4260)$~\cite{4260}, which do not fit well into the conventional
$\ccbar$ picture. 
This has prompted much theoretical activity and proposals
for new models~\cite{new-models}.
Several studies of these states have been performed with the $J/\psi \pi^+\pi^-$ final state~\cite{Jpsipipi}, but no
search using the \ecpp\ final state has been conducted. Such a search
may shed light on the quantum numbers or the internal dynamics of
these states.
In particular, it has been suggested~\cite{Olsen:2004fp} that if the
$X(3872)$ is the $1^1 D_2$ state $\eta_{c2}$, then the branching fraction
$\BR(X(3872)\to\ecpp)$ could be significantly larger than 
$\BR(X(3872)\to J/\psi\pi^+\pi^-)$. 
The quantum numbers $J^{PC}=2^{-+}$ of the
$\eta_{c2}$ are consistent with the results of an angular analysis of
$X(3872)\to J/\psi\pi^+\pi^-$~\cite{Abulencia:2006ma} and would allow production of $X(3872)$ in two-photon fusion.

We present herein a study of the process $\gg \to X \to \ecpp$, 
where $X$ is one of the resonances $\chi_{c2}(1P)$,
$\etac(2S)$, $X(3872)$, $X(3915)$, or $\chi_{c2}(2P)$,
and the $\etac$ is reconstructed in the final state $\kskp$~\cite{charge}.


The data sample was collected with the \babar\ detector
at the PEP-II asymmetric-energy $\epem$ collider located at the SLAC National Accelerator Laboratory. 
It consists of $429.1 \pm 1.9~\invfb$
collected at the energy of the \FourS resonance, 
constituting the entire \babar\ \FourS dataset,
and $44.8 \pm 0.2~\invfb$
collected about 40~\mev\ below the \FourS resonance.
The \babar\ detector is described in detail in Ref.~\cite{babar_nim}.

Samples of Monte Carlo (MC) simulated events are analyzed with the
same reconstruction and analysis procedures as the data sample, following a
$\geant$4-based~\cite{g4} detector simulation~\cite{babar_nim}.
Simulated background samples include $\epem\to q \bar q$ continuum
events $(q=u,d,s,c)$ generated with JETSET~\cite{jetset}, $\Upsilon(4S)\to B \bar B$ decays
generated with EvtGen~\cite{evtgen} and JETSET, and $\epem\to
\tau^+\tau^-$ events generated with KK~\cite{kk2f}. 
In order to study initial-state-radiation (ISR)
background and the invariant-mass resolution,
a sample of $\epem\to\gamma \psi(2S)$ events with 
$\psi(2S)\to J/\psi\pi^+\pi^-$ and $J/\psi\to\kskp$ is generated with
EvtGen. 
The GAMGAM~\cite{gamgam} generator is used to generate signal event
samples for each of the $X$ states studied, with the decay $X\to\ecpp$
simulated with an amplitude that is uniform throughout the decay
phase space, independent of the final-state kinematic variables.
The decay $\etac\to\kskp$ is generated with a uniform amplitude
or with equal and incoherent $K^{*}_{0}(1430)^{-}K^+$ and
$\bar K^{*}_{0}(1430)^{0}  K^0$ contributions.
The GAMGAM generator is also used to generate $\gg\to\etac\to\kskp$ events.


The analysis is performed with two data samples. The sample used to
search for the process $\gg\to X\to\ecpp$ is referred to as the ``main
sample''. Properties of the $\etac$ and its decay into $\kskp$ are
studied with a separate ``control sample'' of $\gg\to\etac\to\kskp$
events. For the main (control) sample, we select events that contain six (four)
charged-particle tracks.

For both samples, charged kaon candidates are identified using
likelihood values calculated from measurements of specific energy loss 
and information from a detector of
internally reflected Cherenkov radiation. 
All other tracks are assumed to be pions. 
A $\KS$ candidate is reconstructed by fitting a $\pi^+\pi^-$ pair to
a common vertex, with invariant mass
in the range $0.491 < m(\pi^+\pi^-) < 0.503~\gevcc$.  A kinematic fit
is performed, constraining $m(\pi^+\pi^-)$ to the
nominal $\KS$ mass~\cite{pdg}.
An $\etac\to\kskp$ decay candidate is reconstructed 
by combining a $\KS$ candidate with a $K^+$ and a $\pi^-$ and requiring the resulting 
invariant mass to lie in the range $2.77 < \metac < 3.22~\gevcc$. 
In the main sample, the decay $X\to\ecpp$ is reconstructed by
combining an $\etac$ candidate with the remaining two tracks in the
event.
A kinematic fit is applied, requiring
the $X$-candidate decay vertex to be consistent with the \epem interaction
region.
The angle $\alpha_{\KS}$ between the $\KS$ momentum vector and the line
connecting the $\etac$ and the $\KS$ decay vertices is required to satisfy
$\cosks>0.99$.

\begin{figure}[htbp!]
\centering
\epsfig{file={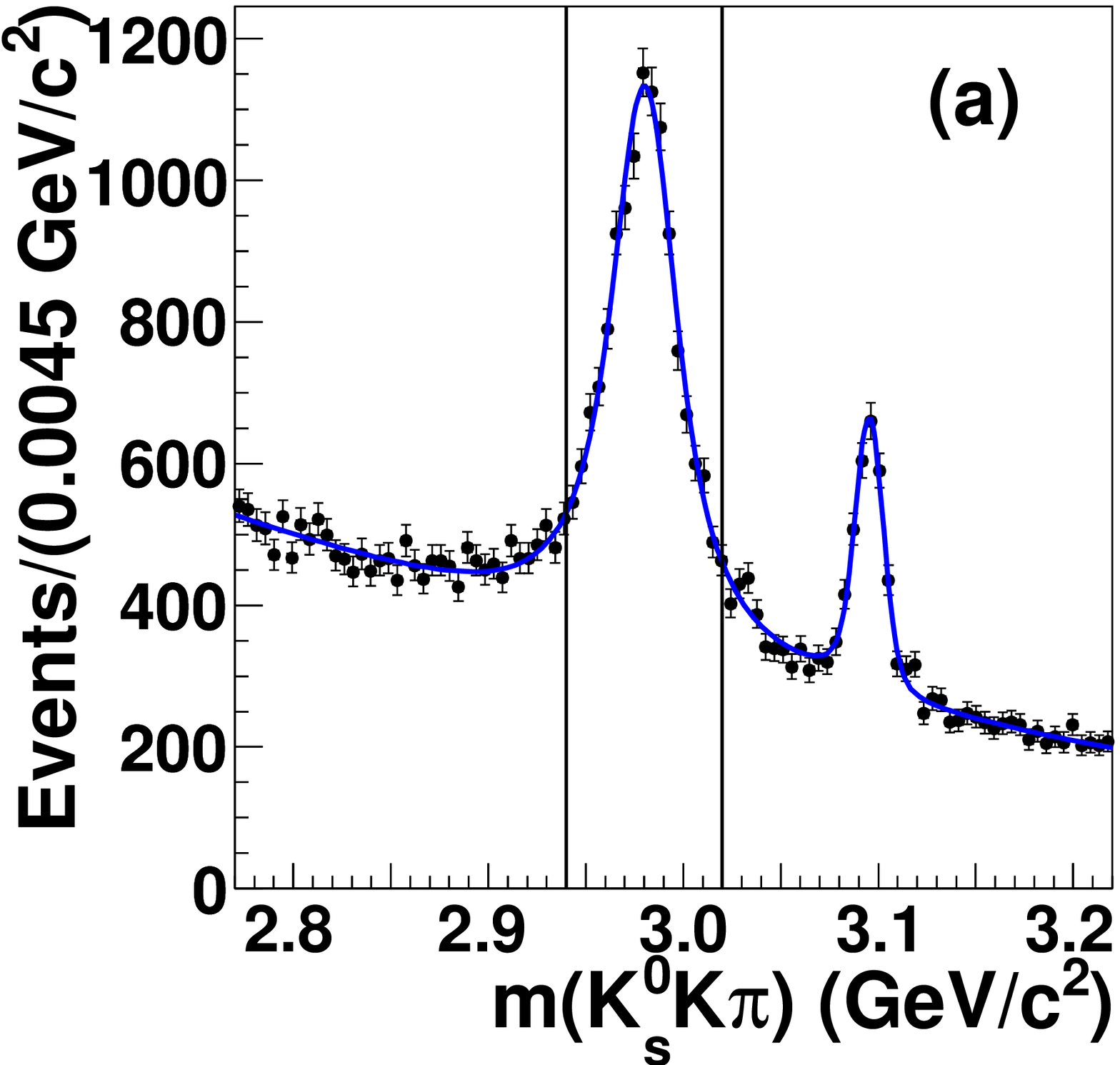},width={1.68in},height={1.68in}}
\epsfig{file={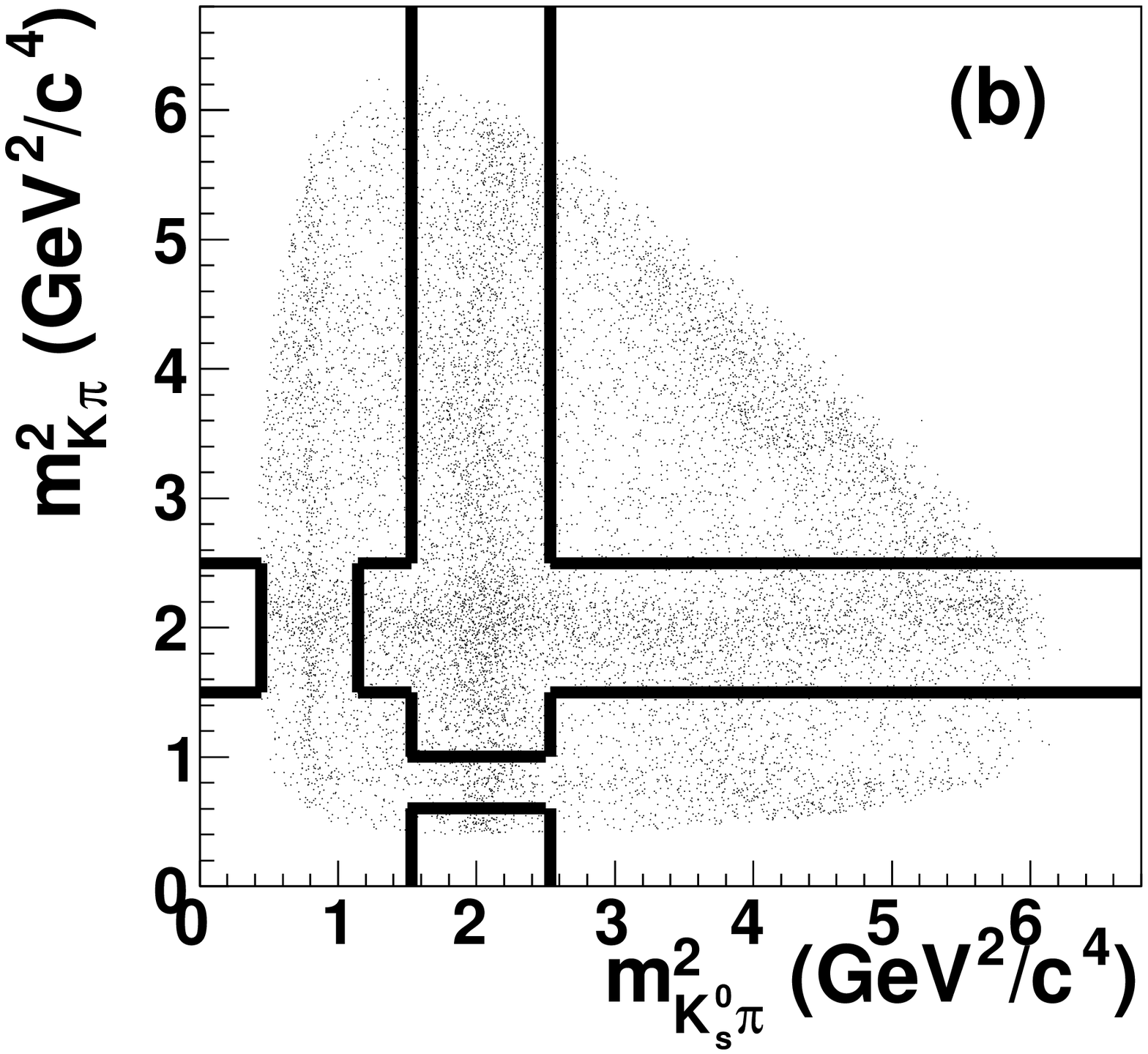},width={1.68in},height={1.68in}}
\epsfig{file={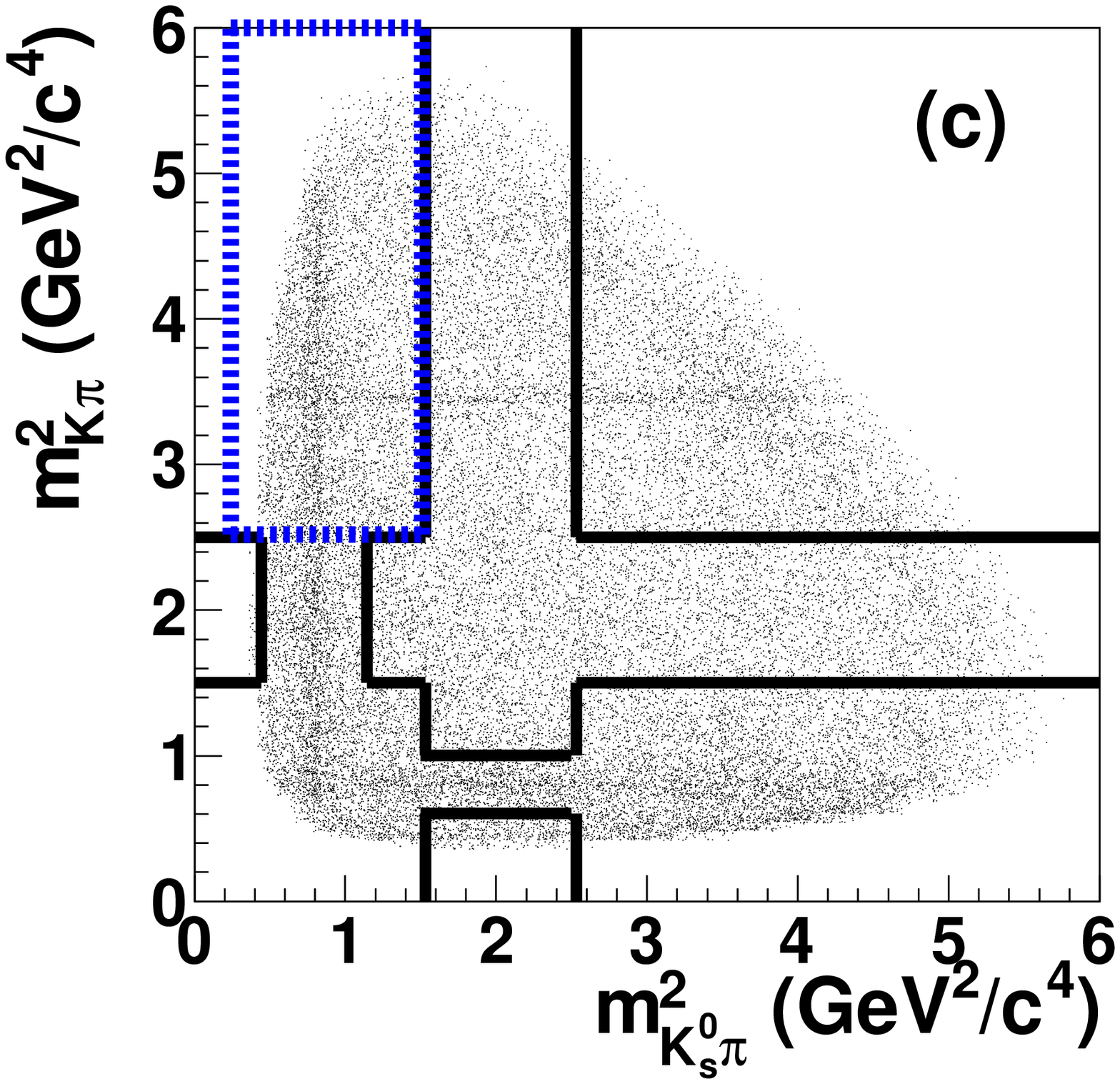},width={1.68in},height={1.68in}}
\epsfig{file={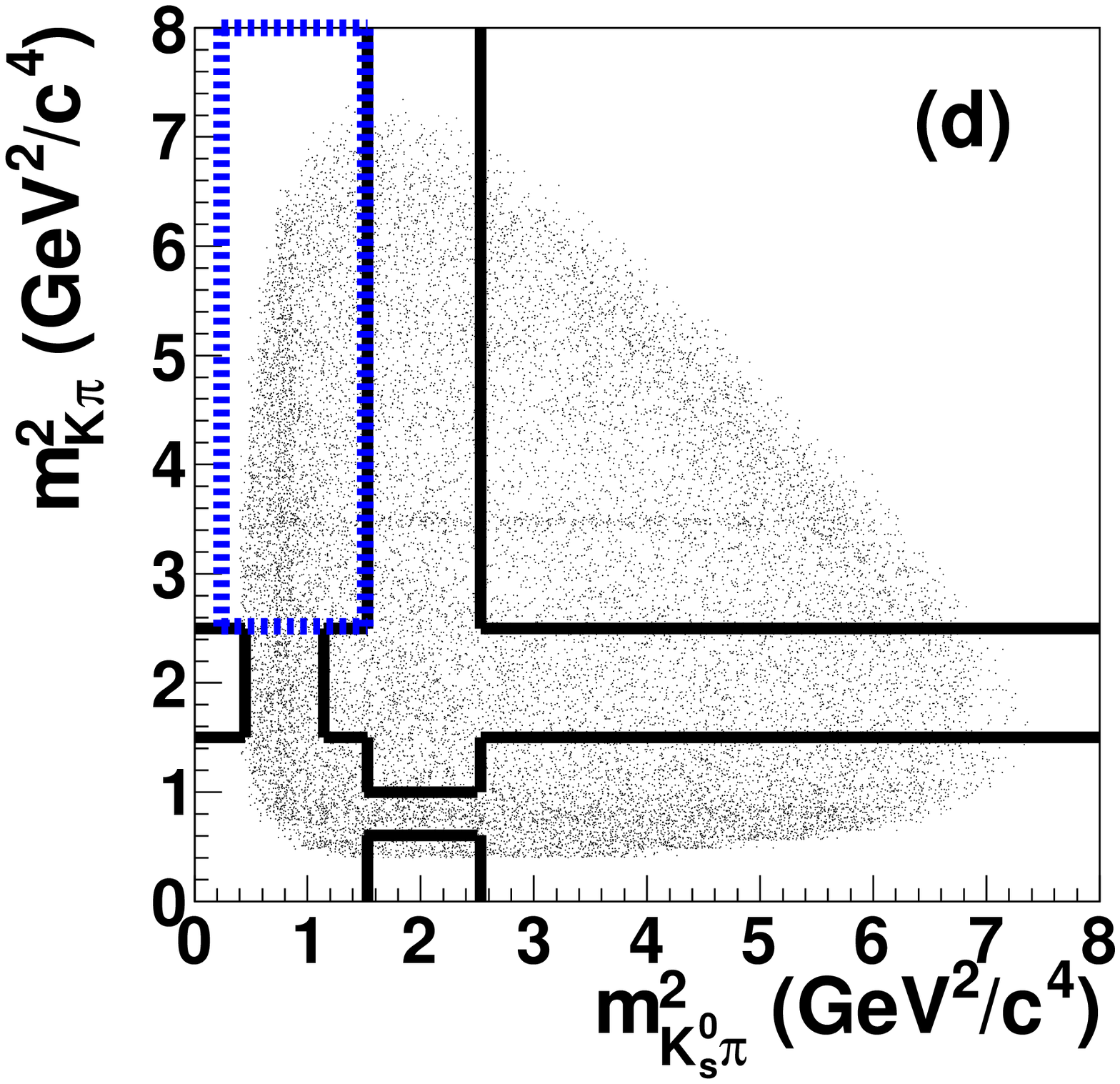},width={1.68in},height={1.68in}}
\epsfig{file={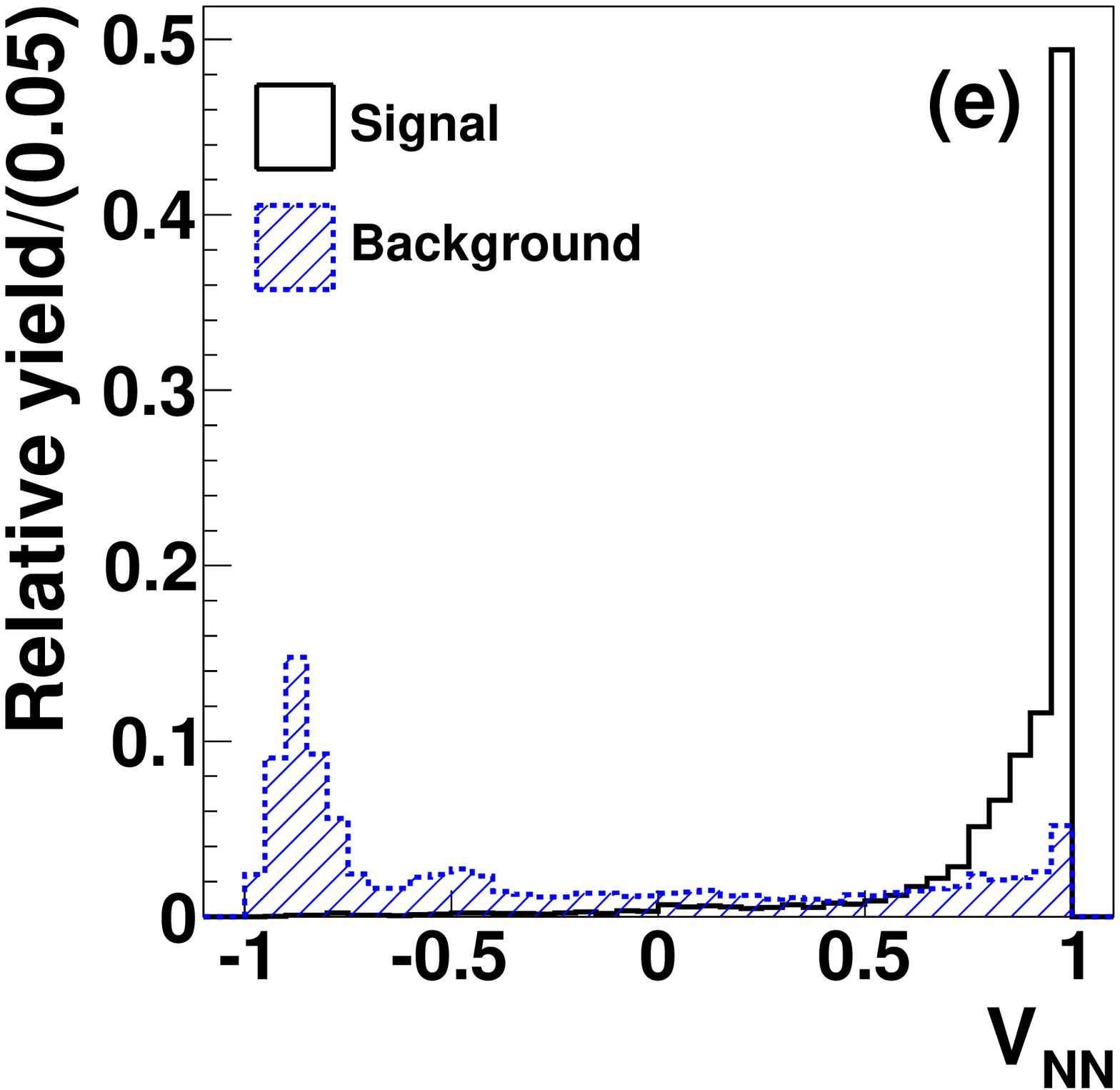},width={1.68in},height={1.68in}}
\epsfig{file={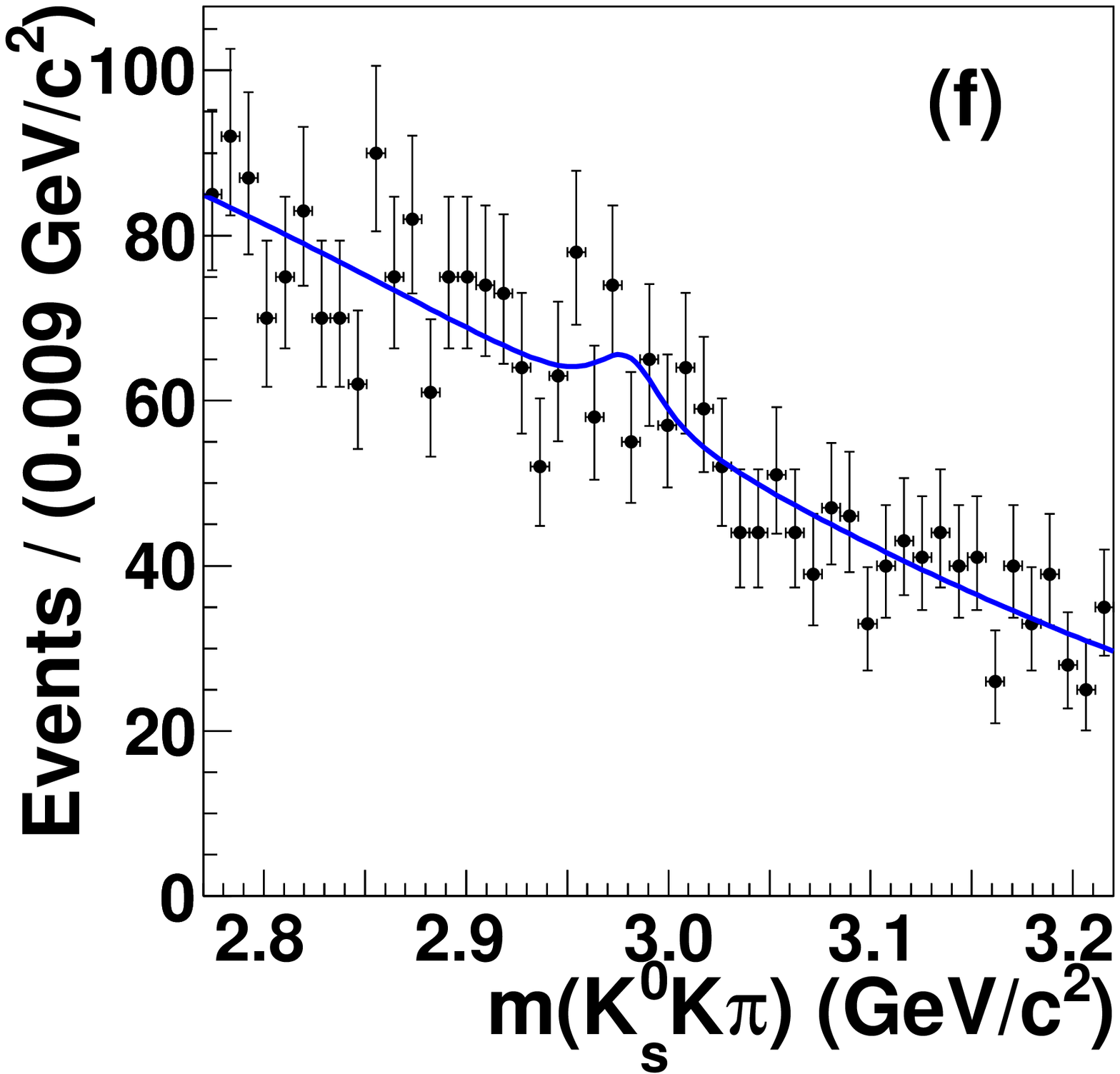},width={1.68in},height={1.68in}}
\caption{\label{fig:DP} (a) The $\metac$ distribution for the control sample.
The vertical lines indicate the \etac\ peak mass region. 
Also shown are the \kskp\ Dalitz-plots for  
(b) control-sample events in the \etac\ peak mass region 
and for main-sample events in the (c) lower and (d) upper 
\etac\ mass sidebands.
Solid black lines indicate the regions defined by the Dalitz-plot selection criteria. 
The dotted blue box in the upper left corner of (c) and (d)
indicates the Dalitz-plot-sideband background region 
used for the neural-network training.
(e) The neural-network output-variable distributions for 
the Dalitz-plot sideband (hatched) and signal MC.
(f) The
result of the step-1 fit (see text).
}
\end{figure}

In the control sample, we require
the polar angle (the angle with respect to the beam axis) $\theta_{\etac}$ of
the $\etac$ candidate to satisfy $ \left| \cosetac \right| > 0.99$ 
and the transverse momentum of the $\etac$
candidate to satisfy $\ptetac< 0.5~\gevc$, both in the center-of-mass (CM) frame of the $\epem$ system. The extra energy in the
event, defined as the total energy in calorimeter clusters not
associated with the identified tracks, is required to satisfy $\Ex <
0.5~\gev$ in the CM frame.  
The $\metac$ distribution of the selected control-sample events, shown 
in Fig.~\ref{fig:DP}(a), exhibits clear $\etac$ and $\jpsi$ peaks,
with the $\jpsi$ produced in ISR events.
In the main sample, continuum background is strongly suppressed 
with the requirements $ \left| \cosx \right| > 0.85$, 
$\ptx< 1.5~\gevc$, and $\Ex <0.8~\gev$, where $\cosx$ and $\ptx$ are the polar angle and
transverse momentum of the $X$ candidate.
In addition, the total visible energy in the event, obtained from all
charged tracks and calorimeter clusters, is required to satisfy
$\evis<10~\gev$ in the laboratory frame.

In the main sample, we suppress QED background by requiring $R2<0.7$,
where $R2$ is the ratio of the second and zeroth Fox-Wolfram
moments~\cite{fox_wolfram}.
We suppress background due to ISR events
with a requirement on the missing mass squared $\mmisssq \equiv
(p_{\epem} - p_X)^2 > 10~\gevccsq$, where $p_{\epem}$
($p_X$) is the total 4-momentum of the beam particles ($X$ candidate).

Additional background suppression in the main sample is obtained by using
the Dalitz plot for the $\etac$ candidates. 
The Dalitz plot is shown in Fig.~\ref{fig:DP}(b) for
control-sample events in the \etac\ peak region $2.94 < \metac <
3.02~\gevcc$, and in Figs.~\ref{fig:DP}(c) and~\ref{fig:DP}(d) for
main-sample events in the lower and upper $\metac$ sidebands $2.8 <
\metac< 2.9~\gevcc$ and $3.05 < \metac <3.2~\gevcc$, respectively.
These distributions indicate that true $\etac\to\kskp$ decays often
proceed via intermediate $K^{*}_{0}(1430)$ states, while background
events contain $K^*(892)$ decays and random combinations.  
Taking advantage of this difference to suppress non-\etac\ background
in the main sample, we require 
$|m^2(\KS\pi^-) - M^2_{K^{*}_{0}(1430)^{-}}| < 0.5~\gevccsq$ or
$|m^2(K^+\pi^-) - M^2_{K^{*}_{0}(1430)^{0}}| < 0.5~\gevccsq$, 
and exclude events that satisfy
$|m^2(\KS\pi^-) - M^2_{K^{*}(892)^{-}}| < 0.35~\gevccsq$ or
$|m^2(K^+\pi^-) - M^2_{K^{*}(892)^{0}}| < 0.2~\gevccsq$,
where $M_R$ is the peak mass of resonance $R$~\cite{pdg}.
The Dalitz-plot region selected by these criteria is enclosed within the
solid lines in Figs.~\ref{fig:DP}(b),~\ref{fig:DP}(c) and~\ref{fig:DP}(d). 
The criteria are the result of
maximizing $\varepsilon^{DP}_{\etac} / \sqrt{\varepsilon^{DP}_{SB}}$, where
$\varepsilon^{DP}_{\etac} = (63.5 \pm 3.2)\%$ is the efficiency of the Dalitz-plot
requirements for \etac\ decays, determined by fitting the \metac\
distribution of the control-sample, and $\varepsilon^{DP}_{SB} = (30.74 \pm 0.21)\%$ is the
corresponding efficiency for main-sample events in the $\metac$
sidebands. 
The $\bar{D}^0\to K^+\pi^-$ band, evident in
Fig.~\ref{fig:DP}(c), becomes insignificant following a neural-network
requirement, described below. Therefore, no explicit effort is
made to remove this source of background.

Further background suppression is achieved by combining six variables
into a neural-network discriminator. Two of the variables are \Ex\
and \ptx. The other four variables, each of which can take on one of five
discrete values, are the outcomes of kaon- and pion-identification
algorithms applied to each of the four charged-particle tracks that are not the daughters of
the $\KS$ candidate. 
The neural network is trained with 
main-sample background events in the Dalitz-plot sideband 
region $m^2(K^+\pi^-) >
2.5~\gevccsq$, $m^2(\KS\pi^-) < 1.5~\gevccsq$, indicated by the dashed
boxes in Figs.~\ref{fig:DP}(c) and~\ref{fig:DP}(d). This region is
chosen since it contains only $(3.40 \pm 0.66)$\% of $\etac$ decays in
the control sample.
We find only insignificant differences in the neural network
signal-to-background separation when using different signal samples
or the mirror Dalitz-plot region $m^2(K^+\pi^-) < 1.5~\gevccsq$,
$m^2(\KS\pi^-) > 2.5~\gevccsq$ for the background.

The distributions of the output-variable $V_{\rm NN}$ are shown in Fig.~\ref{fig:DP}(e).
We find the optimal selection on this variable to be $V_{\rm NN} >
0.84$. 
The efficiency of this selection is $72\%$ for the $\etac(2S)$,
and varies by up to 4\%, depending on the $X$ mass. 
The background efficiency is $(10.4 \pm 0.2)\%$ for the neural-network
training region and $(7.4 \pm 0.2)\%$ for the mirror region.

We find 2863 main-sample events that satisfy all the selection
criteria, with only about 700 events expected from non-$\gg$ 
background MC. We conclude that the majority of the background is 
due to $\gg$ events, for which we have no generic generator. 
More than one $X$ candidate is reconstructed in 3.8\% of the
events. In these cases, we select the candidate for which
$m^2(\KS\pi^-)$ or $m^2(K^+\pi^-)$ is closest to the $K^*_{0}(1430)$ peak.

In addition to these samples, an ISR-produced sample of $\psi(2S)\to
J/\psi\pi^+\pi^-$ events is used to evaluate a systematic
uncertainty associated with the detector resolution. This sample is
selected in the same way as the main sample, except that the
neural-network and Dalitz-plot selections are not applied, the \kskp\
invariant mass is required to be between $3.0$ and $3.2~\gevcc$, and
$\mmisssq$ must be less than $1~\gevccsq$.


We define four categories of events in the main sample: 
signal corresponds to $\gg\to X \to \ecpp$ events;
combinatorial background (\cb), which is by far the most copious background, 
arises from random combinations of final-state particles;
events with a true $\etac\to\kskp$
decay and two pions not originating from an $X$ resonance decay are
categorized as $\etac$-peaking background (\epb);
$X$-peaking background (\xpb) corresponds to decays 
$X\to\kskp\pi^+\pi^-$ that do not proceed through an intermediate $\etac$.

The extraction of the signal yield proceeds in two steps. In step~1, we 
determine the values of the $\metac$-distribution parameters of the combinatorial background
from a one-dimensional fit to $\metac$, without any restrictions
on $\mx$.

In step~2, we extract the signal yield for each $X$ resonance hypothesis from a
two-dimensional fit to the $\metac$ versus \mx\ distribution for events
in an \mx\ window around the resonance peak.
The fits use the unbinned, extended-maximum-likelihood method and are
performed with the RooFit package~\cite{roofit}.

From events in the $\metac$ sidebands, we observe that all
correlation between the \mx\ and \metac\ distributions for the
combinatorial background is accounted for by the phase space
$\phsp(\metac, \mx)$ of the three-body final state consisting of 
the $\pi^+$, $\pi^-$, and the $(\kskp)$ system.
This is used to construct the probability-density function (PDF)~\cite{PDF}
of the step-1 fit. This PDF is a function of $\metac$ with \mx\ as a
conditional variable, and is given by:
\beq 
\E(\metaceq | \mxeq) = N_{\etac} \E_{\etac}(\metaceq) + 
N_{\nonetacsub} \E_{\nonetacsub}(\metaceq | \mxeq),
\label{eq:incl-pdf}
\eeq
where $N_{\etac}$ ($N_{\nonetacsub}$) is the number of events
with (without) a true $\etac\to\kskp$ decay. We have used 
the notation
$\metaceq \equiv \metac$ and $\mxeq \equiv \mx$ for brevity.

The PDF for non-\etac\ events in Eq.~(\ref{eq:incl-pdf}) is 
\beq
\E_{\nonetacsub}(\metaceq | \mxeq) = \P_2(\metaceq; a_1, a_2, \metaceq^0) \phsp(\metaceq , \mxeq),
\label{eq:comb-incl-pdf}
\eeq
where $\P_2(\metaceq; a_1, a_2, \metaceq^0) = 1 + a_1 (\metaceq -
\metaceq^0) + a_2 (\metaceq - \metaceq^0)^2$
is a second-order polynomial and $\metaceq^0 = 3.0~\gevcc$. 
Determination of the coefficients $a_1$, $a_2$ is
the main purpose of the step-1 fit.
The PDF for \etac\ events in Eq.~(\ref{eq:incl-pdf}) is 
$\E_{\etac}(\metaceq) = \W(\metaceq ; M_{\etac}, \Gamma_{\etac}, \vec r_{\metaceq})$,
where $\W$ is a relativistic Breit-Wigner function 
$\left[(\tilde{m}_{3}^2 - M_{\etac}^2)^2 + M_{\etac}^2 \Gamma_{\etac}^2\right]^{-1}$
convolved with a detector resolution function 
$\R(\metaceq - \tilde{m}_{3} ; \vec r_{\metaceq})$ 
that depends on a set of parameters $\vec r_{\metaceq}$ and the 
true invariant mass $\tilde{m}_{3}$ of the $\kskp$ system. 
The resolution function is the sum of two Crystal Ball
functions~\cite{Skwarnicki:1986xj} with oppositely-directed tails and
common Gaussian-parameter values. The resolution-function parameter
values are determined from a fit to the MC.

In addition to $a_1$ and $a_2$, the parameter values determined in 
the step-1 fit are the yields $N_{\etac}$ and $N_{\nonetacsub}$, and the
mass $M_{\etac}$ and width $\Gamma_{\etac}$ of the \etac\ peak.
In order to obtain $M_{\etac}$ and $\Gamma_{\etac}$ from the data,
the step-1 fit is performed simultaneously for the main sample and 
the control sample. The PDF for the control sample is 
\beqa
\E'(\metaceq) &=& 
   N'_{J/\psi} \W(\metaceq ; M_{J/\psi}, \Gamma_{J\psi}, \vec r_{\metaceq}) 
    \\
            &+& N'_{\etac} \E_{\etac}(\metaceq) +  N'_{\rm bgd} \P_2(\metaceq ; a'_1, a'_2, \metaceq^0). \nonumber
\eeqa
Additional control-sample
parameter values determined in the fit are the peak $J/\psi$ mass
$M_{J/\psi}$, the background parameters $a'_1$, $a'_2$, and the
\etac, $J/\psi$, and background event yields $N'_{\etac}$,
$N'_{J\psi}$, and $N'_{\rm bgd}$.

The \metac\ distribution of the data and the step-1 PDF are
shown in Fig.~\ref{fig:DP}(f).
The fitted parameter values are
$a_1 = 1.24 \pm 0.19$ $(\gevmass)^{-1}$,
$a_2 = 0.2 \pm 1.4$ $(\gevmass)^{-2}$,
$N_{\etac} = 50 \pm 37$,
$N'_{\etac} = 10350 \pm 300$, and 
$N'_{J/\psi} = 1877 \pm 90$.
The large relative uncertainties for $a_1$ and $a_2$ are the result of 
the near linearity of the $\metac$ distribution and the correlation between
the two parameters, 
which is taken into account in the evaluation of systematic uncertainties.
The $\etac$ parameter values determined in the step-1 fit are
$\Gamma_{\etac} =31.7 \pm 1.5~\mevcc$ and
$M_{\etac} = 2.98285 \pm 0.00038~\gevcc$, where the
uncertainties are statistical only. 
These results are consistent with previous measurements~\cite{pdg}.

The PDF for the step-2 fit is a linear combination 
of the PDFs of the four event types, 
\beq
\P =  N_{\rm sig} \P_{\rm sig}  + N_{\cb}  \P_{\cb}  
   + N_{\epb} \P_{\epb} + N_{\xpb} \P_{\xpb}. 
\label{eq:excl-pdf}
\eeq
The signal PDF is a relativistic Breit-Wigner function convolved with the
resolution function, for both $\metaceq$ and $\mxeq$:
\beq
\P_{\rm sig}(\metaceq, \mxeq) = \E_{\etac}(\metaceq)  \,
          \W(\mxeq ; M_X, \Gamma_X, \vec r_{\mxeq}),
\label{eq:excl-sig-pdf}
\eeq
where $M_X$ and $\Gamma_X$ are the known mass and width of the
resonance of interest~\cite{pdg,delAmoSanchez:2011bt,Uehara:2009tx},
and $\vec r_{\mxeq}$ are the parameters of the \mx\
resolution function $\R(\mxeq - \tilde{m}_{5} ; \vec r_{\mxeq})$, obtained from a fit to signal MC.
The combinatorial-background PDF is
\beq
\P_{\cb}(\metaceq, \mxeq) = \E_{\nonetacsub}(\metaceq | \mxeq)  \,
      \C_2(\mxeq ; b_1^{\cb}, b_2^{\cb}),
\label{eq:excl-comb-pdf}
\eeq
where $\C_2(\mxeq ; b_1^{\cb}, b_2^{\cb})$ is a second-order
Chebychev polynomial with first- (second-) order coefficients
$b_1^{\cb}$ ($b_2^{\cb}$).
The $\etac$-peaking background PDF is 
\beq
\P_{\epb}(\metaceq, \mxeq) = \E_{\etac}(\metaceq)  \, 
         \C_1(\mxeq ; b_1^{\epb}),
\label{eq:excl-etac-peak-pdf}
\eeq
where $\C_1(\mxeq ; b_1^{\epb})$ is a first-order Chebychev polynomial.
The $X$-peaking background PDF is 
\beqa
\P_{\xpb}(\metaceq, \mxeq) &=& \P_1(\metaceq ; c_1^{\xpb}, \metaceq^0)  \\
                        & & \W(\mxeq ; M_X, \Gamma_X, \vec r_{\mxeq}), \nonumber
\label{eq:excl-X-peak-pdf}
\eeqa
where $\P_1(\metaceq ; c_1^{\xpb}, \metaceq^0)$ is a first-order polynomial.
The parameter values determined with the step-2 fit are the four yields of
Eq.~(\ref{eq:excl-pdf}) and the background shape parameters
$b_1^{\cb}$, $b_2^{\cb}$, $b_1^{\epb}$, and $c_1^{\xpb}$.

\begin{figure}[!htbp]
\centering
\epsfig{file={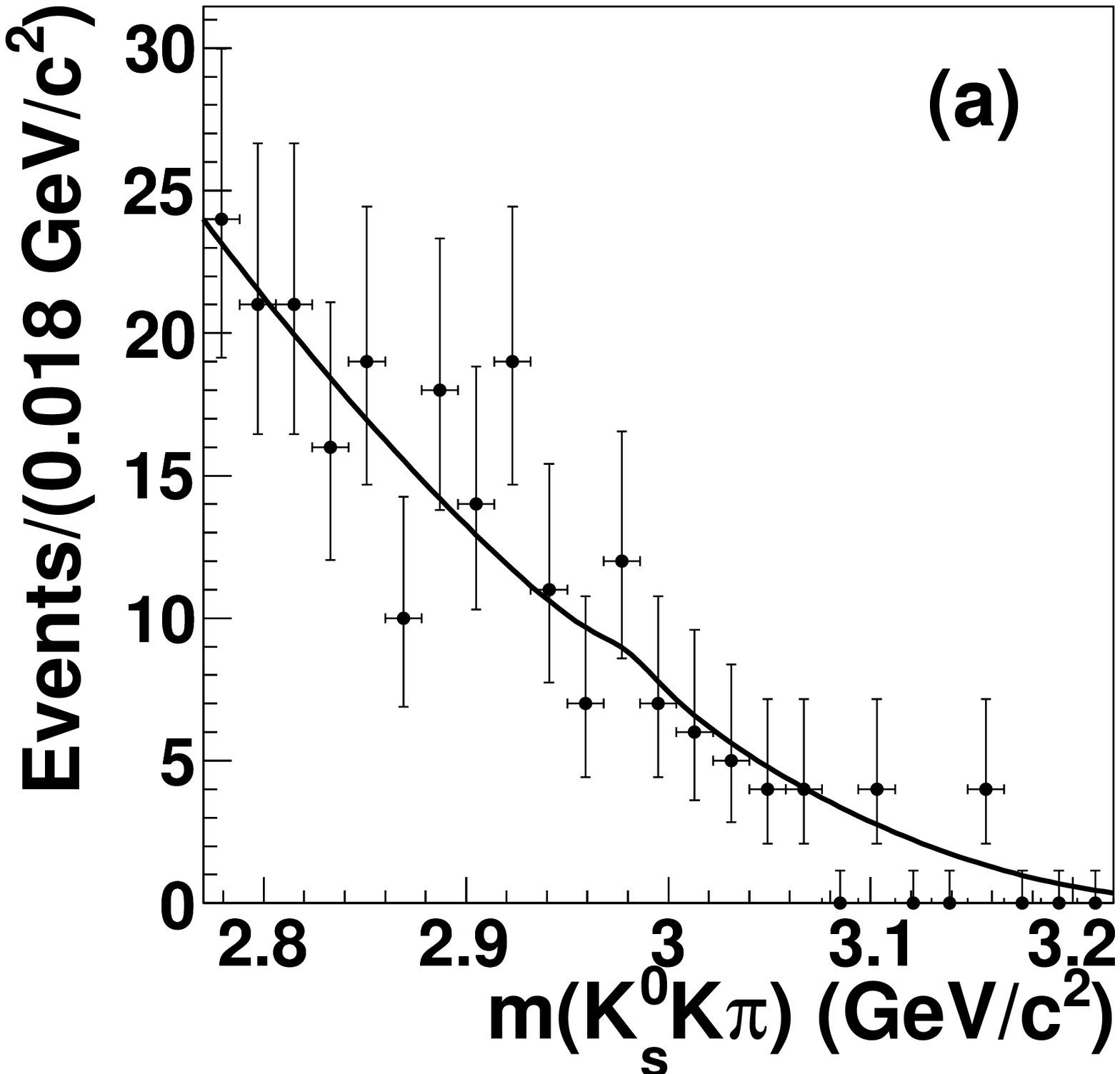},width={1.68in},height={1.68in}}
\epsfig{file={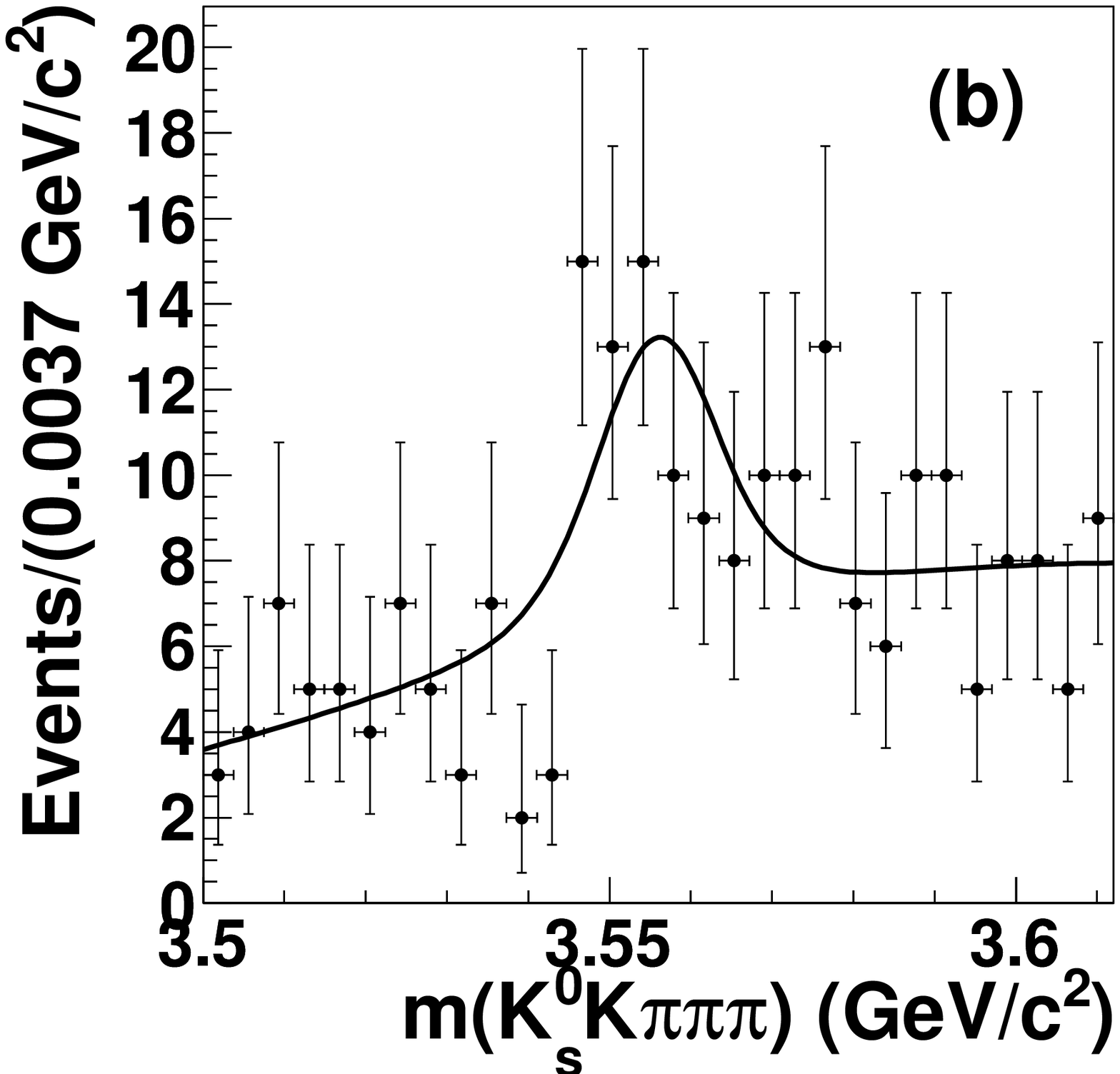},width={1.68in},height={1.68in}} \\
\epsfig{file={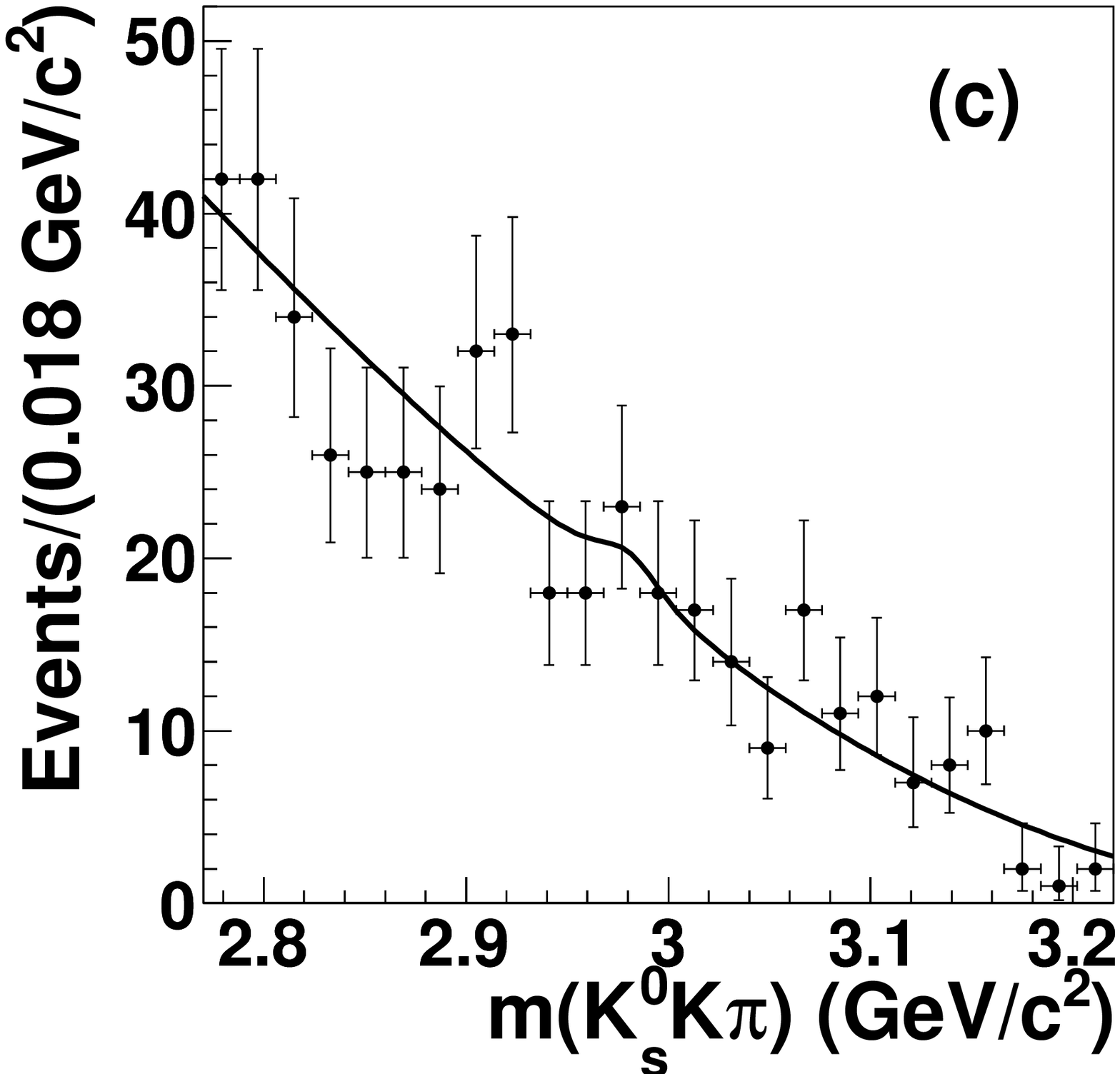},width={1.68in},height={1.68in}} 
\epsfig{file={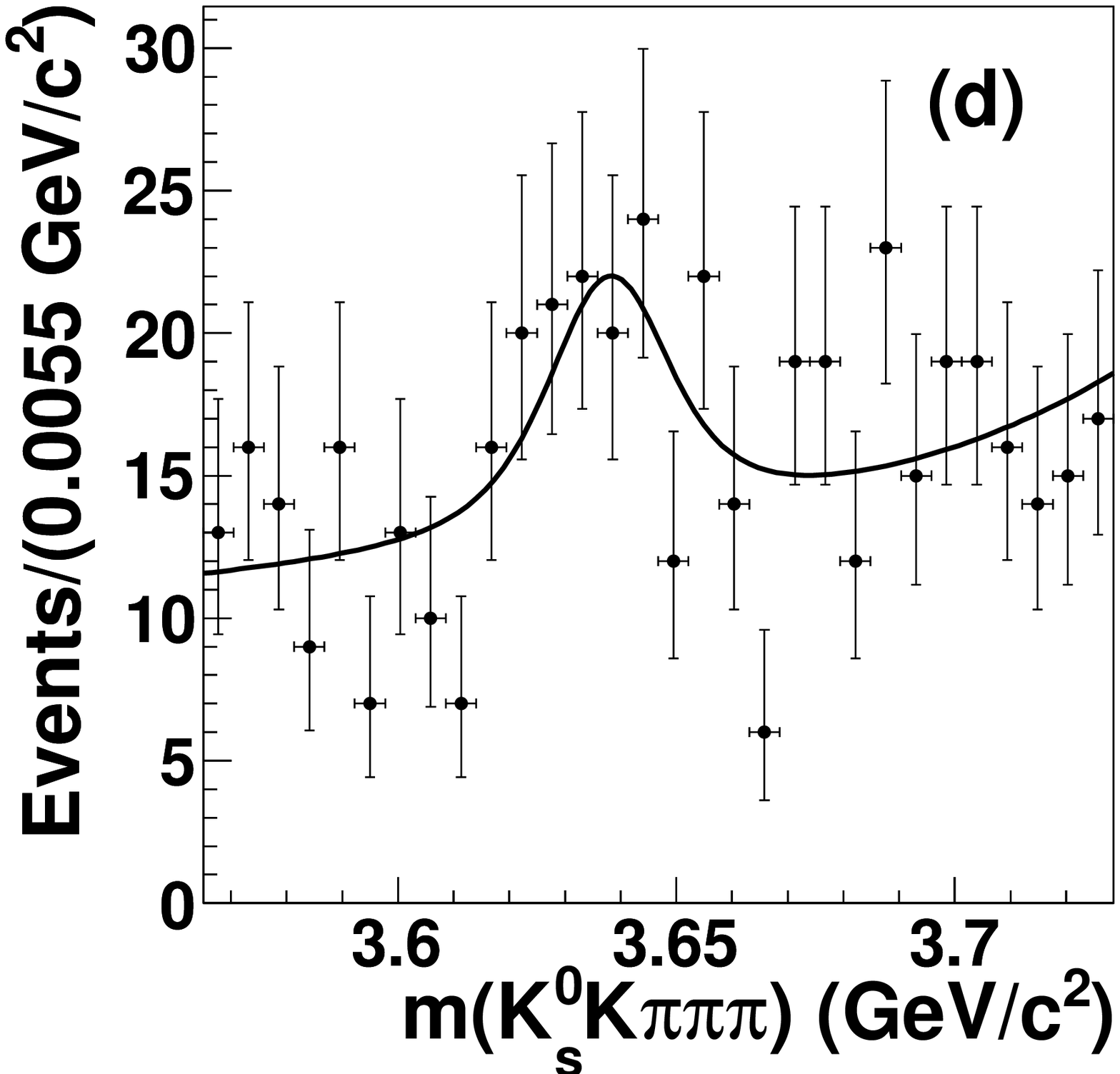},width={1.68in},height={1.68in}} \\
\epsfig{file={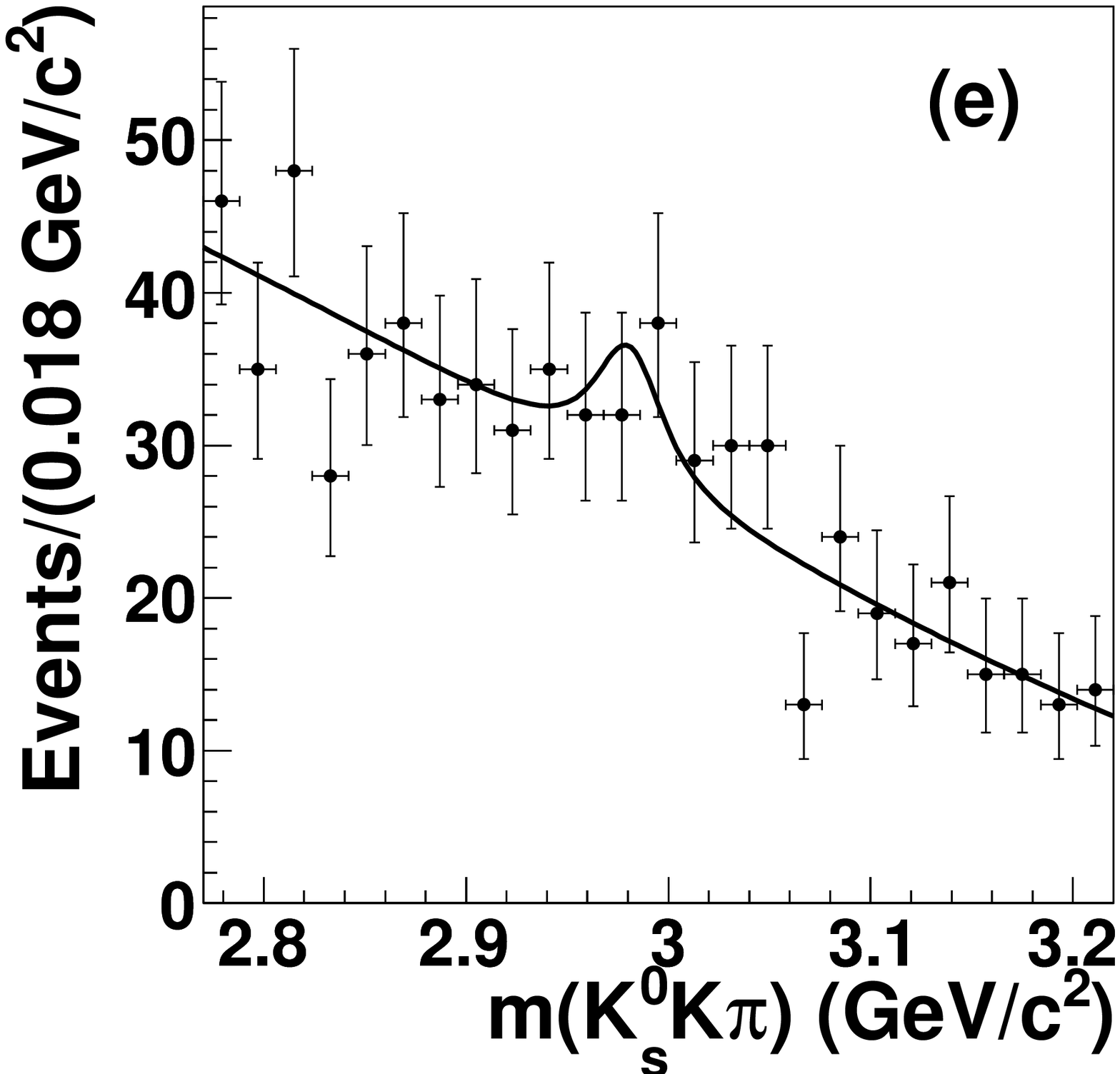},width={1.68in},height={1.68in}}
\epsfig{file={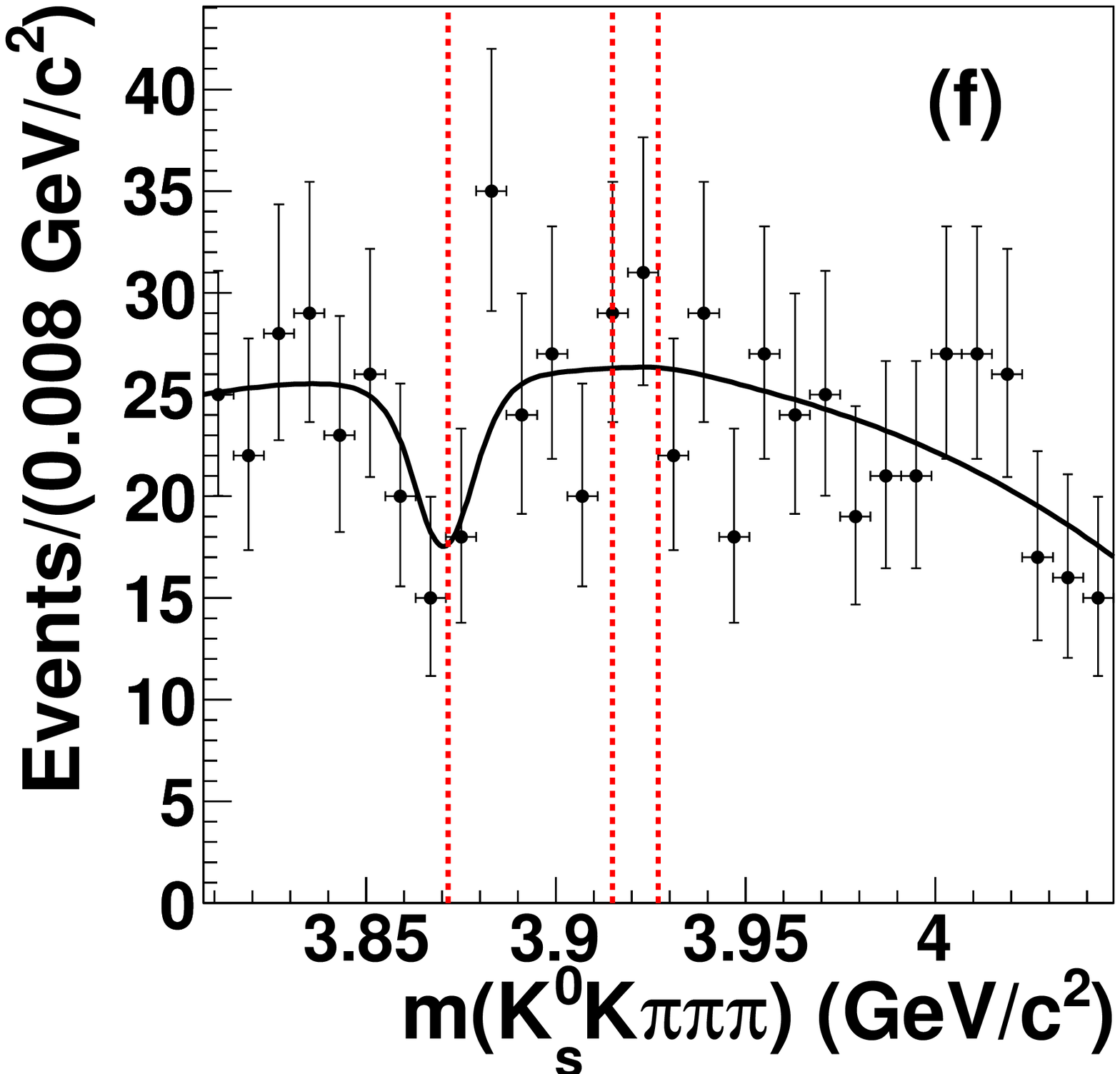},width={1.68in},height={1.68in}}
\caption{\label{fig:results} Distributions of  (a,c,e) 
$\metac$ and (b,d,f) $\mx$ 
with the step-2 fit PDF overlaid for the fit regions of the 
(a,b) $\chi_{c2}(1P)$, 
(c,d) $\etac(2S)$, 
and (e,f) $X(3872)$,  $X(3915)$ and $\chi_{c2}(2P)$.
The vertical dashed lines in (f) indicate the peak mass
positions of the $X(3872)$, $X(3915)$, and $\chi_{c2}(2P)$~\cite{pdg}.  }
\end{figure}

The step-2 fit is performed four times in different $\mx$ windows, 
fitting for the
(1) $\chi_{c2}(1P)$, (2) $\etac(2S)$, (3) $X(3872)$ and $X(3915)$, 
or (4) $X(3872)$ and $\chi_{c2}(2P)$ resonances.
A simultaneous fit to the three resonances $X(3872)$, $X(3915)$, and
$\chi_{c2}(2P)$ is observed to be unstable when tested with parametrized
MC experiments, due to the large number of fit parameters, small signal, and large
overlap of the $X(3915)$ and $\chi_{c2}(2P)$ lineshapes. Therefore, we
conduct fits (3) and (4) separately to test for the existence of a signal
for  either set of lineshape parameters.
The \mx\ and \metac\ distributions and fit functions are shown in
Fig.~\ref{fig:results}. The difference between the fit function of 
fit (3) and that of fit (4) is almost indistinguishable within the thickness of the curve in 
Fig.~\ref{fig:results}(f).
The fitted signal yields are
summarized in Table~\ref{tab:results}.

No significant signal or peaking background is observed in any of the
fits. However, a hint of $X$-peaking background is visible in the
$\chi_{c2}(1P)$ and $\etac(2S)$ fits of Figs.~\ref{fig:results}(b) and
(d), with event yields of $33 \pm 14$ and $47 \pm 24$, respectively,
where the uncertainties are statistical only.
This may be due to decays of $\chi_{c2}(1P)$ and
$\etac(2S)$ into $\kskp \pi^+\pi^-$~\cite{Nakazawa:2010zz}, 
which are suppressed in this analysis by the $2.77 < \metac < 3.22$~\gevcc requirement.
Fits (3) and (4) yield insignificant X-peaking background, roughly canceling the negative signal yields.
The results shown in Table~\ref{tab:results} for the $X(3872)$ are
obtained from fit (4). The $X(3872)$ yield from the 
$X(3915)$ fit is 1.6 events lower.
Since no signal is observed for the $X(3915)$ or the $\chi_{c2}(2P)$, we
obtain a conservative upper limit on the yield of the $X(3915)$ ($\chi_{c2}(2P)$) by
fixing the $\chi_{c2}(2P)$ ($X(3915)$) yield to zero.
%


We estimate systematic uncertainties on the signal yields associated
with the fit procedure by repeating the fits with the variations described
below and adding the different uncertainties in quadrature.
We account for uncertainties in the $X$ mass and width values by
varying them within their uncertainties~\cite{pdg}. This is the source
of the largest signal-yield systematic uncertainty, except for the 
$\chi_{c2}(1P)$.
The order of the polynomial in each PDF is varied to
account for uncertainties due to background modeling. 
We vary the resolution-function Gaussian width by $2~\mevcc$ for the
\metac\ PDF to account for a difference between the $J/\psi$ width in
MC and in the control sample, and by $0.9~\mevcc$ for the \mx\ PDF
due to a difference in the $\psi(2S)$ width between MC and data. 
An additional uncertainty is evaluated by using 
the sum of three Gaussians to define the resolution function.
To address the possibility that correlations between the \mxeq\ and
\metaceq\ distributions are not taken fully into account by the phase-space
factor $\phsp(\metaceq , \mxeq)$ in Eq.~(\ref{eq:comb-incl-pdf}), we replace the parameters $a_i$ of
Eq.~(\ref{eq:comb-incl-pdf}) by $a_i (1 + a''_i \mxeq)$. The values of the parameters
$a''_i$ are found to be consistent with zero, and we conservatively use 
their uncertainties to evaluate the systematic uncertainty on the 
signal yield.
The effect of not accounting for phase-space correlations between \metaceq\ and \mxeq\ in the 
signal and $\etac$-peaking background PDFs is determined to be
small compared to other systematic uncertainties,
except for the $\chi_{c2}(1P)$, for which this uncertainty is dominant 
and equals 2.4 events.
Statistical uncertainties from the step-1 fit are propagated to the 
step-2 fit, accounting for correlations among the parameters.

We test the entire fit procedure using parameterized MC experiments
generated with the PDFs of Eqs.~(\ref{eq:incl-pdf}) and~(\ref{eq:excl-pdf}).
A bias of up to two events on the signal yield is found and used as a
correction that is accounted for in the values shown in Table~\ref{tab:results}.
A systematic uncertainty on this correction is evaluated by repeating 
this study after varying the generated signal yield by its
statistical uncertainty in the data fit.

Since the background is mostly combinatorial, we do not expect
significant interference between signal and background. In addition, the
small signal yields make the evaluation of such interference effects
unreliable. Therefore, we do not attempt to account for possible
interference.

We evaluate systematic uncertainties on track and $\KS$ reconstruction
efficiencies, accounting for the momentum and angular
distribution of signal tracks, as well as on the 
uncertainty of the Dalitz-plot requirement efficiency.
A 2\% systematic uncertainty is assigned due to differences between
the distributions of the selection variables in the control
sample and in $\gg\to\etac\to\kskp$ MC. 
Differences between the data and MC distributions of the
particle-identification variables are studied using a high-purity sample of
$D^{*+} \to \pi^+ D^0$, $D^0 \to K^- \pi^+$ events, and found to have
negligible impact on the efficiency.

We account for uncertainties in the $X\to\ecpp$ decay amplitude, which is
uniform in our simulated signal samples, by weighting events according to 
$\left(m^2(\pi\pi) - 4 M_\pi^2\right)^2$~\cite{Voloshin:2002xh}, where
$m^2(\pi\pi)$ is the squared dipion mass and $M_\pi$ is the $\pi^-$ mass.
From the weighted sample, we extract an efficiency correction of up to 4.6\%
(incorporated into the values in Table~\ref{tab:results}) 
and a systematic uncertainty of the same magnitude.  

Finally, we account for a 0.45\% uncertainty on the
integrated luminosity, for the uncertainties on the $\KS$, $\etac$, and
\etacp\ branching fractions~\cite{pdg}, and for MC-statistical uncertainties.


The results are summarized in Table~\ref{tab:results}.
From the signal yield $N_{\rm sig}$ of each resonance, the integrated
luminosity $L$, and the signal efficiency $\varepsilon$, we compute
the product $\sigma\BR = N_{\rm sig} /(L\varepsilon)$ of the $\epem\to
X\epem$ production cross section and the $X\to\ecpp$ branching
fraction. 
We also evaluate the results in terms of the product
$\Gamma_{\gg}\BR$, where $\Gamma_{\gg}$ is the two-photon width of the
resonance, by utilizing the GAMGAM generator to determine the cross
section as a function of $\Gamma_{\gg}$.  A 10\% uncertainty is
assigned to the GAMGAM calculation~\cite{GGtoZtoDD}.
Since we find no significant signal for the $X$ resonances,
we calculate 90\% confidence-level (CL) Bayesian upper limits on these quantities, 
assuming a Gaussian likelihood incorporating statistical and systematic uncertainties.

Using the efficiency-corrected yields for the $\chi_{c2}$ and $\etac(2S)$ 
from~\cite{delAmoSanchez:2011bt} and the branching
fractions of their decays into $\kskp$, we find the relative
branching fractions
\beqa\label{eq:etacRBF}
\frac{{\cal B}(\etac(2S)\to \eta_{c} \pi^{+} \pi^{-})}{{\cal B}(\etac(2S)\to \KS K^+ \pi^-)} &=& 4.9^{+3.5}_{-3.3} \pm 1.3 \pm 0.8, 
\\
\frac{{\cal B}(\chi_{c2}(1P)\to \eta_{c} \pi^{+} \pi^{-})}{{\cal B}(\chi_{c2}(1P)\to \KS K^+ \pi^-)} &=& 14.5^{+10.9}_{-8.9}  \pm 7.3 \pm 2.5, \nonumber
\eeqa
where the first uncertainty is statistical, the second is systematic, 
and the third is due to the uncertainty on ${\cal B}(\etac \to \kskp)$~\cite{pdg}.
The 90\% CL upper limits on the two ratios in Eqs.~(\ref{eq:etacRBF})
are $10.0$ and $32.9$, respectively.
Using ${\cal B}(\etac(2S)\to \KS K^+ \pi^-)$ and ${\cal
B}(\chi_{c2}(1P)\to \KS K^+ \pi^-)$ from Ref.~\cite{pdg}, we obtain
the 90\% CL upper limits ${\cal B}(\etac(2S)\to \eta_{c} \pi^{+}
\pi^{-}) < 7.4\%$ and ${\cal B}(\chi_{c2}(1P)\to \eta_{c} \pi^{+}
\pi^{-})<2.2\%$.
\begin{table*}[!htb]
\centering
\caption{\label{tab:results} Results of the step-2 fits. For each
resonance $X$, we show the peak mass and width used in the PDF (from Refs.~\cite{pdg,delAmoSanchez:2011bt,Uehara:2009tx});
the mass range of the fit; the efficiency; the
bias-corrected signal yield with statistical and systematic uncertainties; 
the product of the $\gg\to X$
production cross section and $X\to\ecpp$ branching fraction, and the
90\% CL upper limit (UL) on this product; the product of the two-photon
partial width $\Gamma_{\gg}$ and the $X\to\ecpp$ branching fraction,
and the 90\% CL upper limit on this product. For the $X(3872)$ and the $X(3915)$ we
assume $J=2$.}
\begin{tabular}{|l|c|c|c|c|c|c|c|c|c|}
\hline\hline
\multirow{2}{*}{Resonance} &\multirow{2}{*}{$M_X$ (\mevcc)} & \multirow{2}{*}{$\Gamma_X$ (\mev)} & \multirow{2}{*}{$\mxeq$ Range (\gevcc)} &       \multirow{2}{*}{ $\varepsilon$ (\%)}    &\multirow{2}{*}{$N_{\rm sig}$}    &  \multicolumn{2}{c|}{$\sigma \BR$(fb)}  & \multicolumn{2}{c|}{$\Gamma_{\gg}$\BR(eV)} \\
\cline{7-10}
& & & & & & Central value & UL & Central value & UL\\
\hline
$\chi_{c2}(1P)$ & $3556.20 \pm 0.09$& $1.97 \pm 0.11 $&  3.500-3.612   &  $ 3.60  \pm 0.39 $ & $10.2^{+7.7}_{-6.3} \pm 3.5$  &  $37^{+28}_{-23}\pm 15$    & $80$        &  $7.2^{+5.5}_{-4.4} \pm 2.9$    &  $15.7$  \\
\etacp          & $3638.5 \pm 1.7$ & $13.4 \pm 5.6$ &  3.565-3.728   &  $ 3.53  \pm 0.35 $ & $17^{+12}_{-11} \pm 3$ &  $61^{+44}_{-41} \pm 16$    & $123$        & $65^{+47}_{-44} \pm 18$       &  $133$ \\
$X(3872)$       & $3871.57 \pm 0.25$ & $ 3.0 \pm 2.1 $ &  3.807-4.047   &  $ 3.92  \pm 0.38 $ & $ -4.7^{+7.9}_{-6.9}  \pm 2.8$ &  $ -16^{+26}_{-23} \pm 10$   & $38$         & $-4.5^{+7.7}_{-6.7} \pm 2.9$   &  $11.1$  \\
$X(3915)$       & $ 3915.0 \pm 3.6 $ & $17.0 \pm 10.4 $ &  3.807-4.047   &  $ 3.79  \pm 0.37 $ & $ -13^{+11}_{-11}\pm 7$ &  $-44^{+38}_{-38} \pm 25$   & $53$         & $-13^{+12}_{-12} \pm 8$ &  $16$ \\
$\chi_{c2}(2P)$ & $3927.2 \pm 2.6$ & $24 \pm 6$  &  3.807-4.047   &  $ 3.75  \pm 0.36 $ & $-15^{+14}_{-13}\pm 4$ &  $-53^{+49}_{-46} \pm 18$   & $60$         & $-16^{+15}_{-14} \pm 6$ &  $19$ \\
\hline\hline
\end{tabular}
\end{table*}



In summary, we report a study of the process $\gg \to \etac \pi^+ \pi^-$
and provide, for the first time, upper limits on the branching fractions of $\chi_{c2}(1P)$ and 
$\etac(2S)$ decays to $ \etac \pi^+\pi^-$ relative to the branching fractions of the decays into $\KS K^+\pi^-$. 
We also report upper limits on the products $\sigma \BR$ and $\Gamma_{\gg}\BR$ for
the $\chi_{c2}(1P)$, $\etac(2S)$, $X(3872)$, $X(3915)$, and $\chi_{c2}(2P)$ resonances. \\

\indent We are grateful for the 
extraordinary contributions of our \pep2\ colleagues in
achieving the excellent luminosity and machine conditions
that have made this work possible.
The success of this project also relies critically on the 
expertise and dedication of the computing organizations that 
support \babar.
The collaborating institutions wish to thank 
SLAC for its support and the kind hospitality extended to them. 
This work is supported by the
US Department of Energy
and National Science Foundation, the
Natural Sciences and Engineering Research Council (Canada),
the Commissariat \`a l'Energie Atomique and
Institut National de Physique Nucl\'eaire et de Physique des Particules
(France), the
Bundesministerium f\"ur Bildung und Forschung and
Deutsche Forschungsgemeinschaft
(Germany), the
Istituto Nazionale di Fisica Nucleare (Italy),
the Foundation for Fundamental Research on Matter (The Netherlands),
the Research Council of Norway, the
Ministry of Education and Science of the Russian Federation, 
Ministerio de Ciencia e Innovaci\'on (Spain), and the
Science and Technology Facilities Council (United Kingdom).
Individuals have received support from 
the Marie-Curie IEF program (European Union) and the A. P. Sloan Foundation (USA).


\begin{thebibliography}{99}
\bibitem{charge}
The use of charge conjugate reactions is implied throughout this paper.
\bibitem{PDF}
All PDFs include implicit normalization coefficients that are not mentioned.
\end{thebibliography}
\end{document}